\documentclass{article}
\usepackage{ijcai23}

\usepackage{times}
\usepackage{soul}
\usepackage{url}
\usepackage[hidelinks]{hyperref}
\usepackage[utf8]{inputenc}
\usepackage[small]{caption}
\usepackage{graphicx}
\usepackage{amsmath}
\usepackage{amsthm}
\usepackage{booktabs}
\usepackage{algorithm}
\usepackage[switch]{lineno}
\usepackage{caption}
\usepackage{subcaption}
\usepackage[raggedrightboxes]{ragged2e}
\usepackage{tabularx}
\usepackage{arydshln}
\usepackage{xcolor}
\usepackage{algpseudocode}
\usepackage{blindtext}
\usepackage{listings}
\usepackage{color}
\usepackage{lipsum}  
\definecolor{dkgreen}{rgb}{0,0.6,0}
\definecolor{gray}{rgb}{0.5,0.5,0.5}
\definecolor{mauve}{rgb}{0.58,0,0.82}

\title{Copilot for Xcode: Exploring AI-Assisted Programming by Prompting Cloud-based Large Language Models}

\author{
Chee Wei Tan$^1$
\and
Shangxin Guo$^2$\and
Man Fai Wong$^2$\And
Ching Nam Hang$^2$
\affiliations
$^1$Nanyang Technological University\\
$^2$City University of Hong Kong\\
\emails
cheewei.tan@ntu.edu.sg,
\{sxguo2-c,mfwong29-c,cnhang3-c\}@my.cityu.edu.hk
}
\begin{document}

\maketitle

\begin{abstract}
This paper presents an AI-assisted programming tool called {\it Copilot for Xcode} for program composition and design to support human software developers. By seamlessly integrating cloud-based Large Language Models (LLM) with Apple's local development environment, Xcode, this tool enhances productivity and unleashes creativity for software development in Apple software ecosystem (e.g., iOS apps, macOS). Leveraging advanced natural language processing (NLP) techniques, {\it Copilot for Xcode} effectively processes source code tokens and patterns within code repositories, enabling features such as code generation, autocompletion, documentation, and error detection. Software developers can also query and make ``small" decisions for program composition, some of which can be made simultaneously, and this is facilitated through prompt engineering in a chat interface of {\it Copilot for Xcode}. Finally, we present simple case studies as evidence of the effectiveness of utilizing NLP in Xcode to prompt popular LLM services like OpenAI ChatGPT for program composition and design.
\end{abstract}

\begin{figure*}[!t]
\centering
\includegraphics[width=0.6\linewidth]{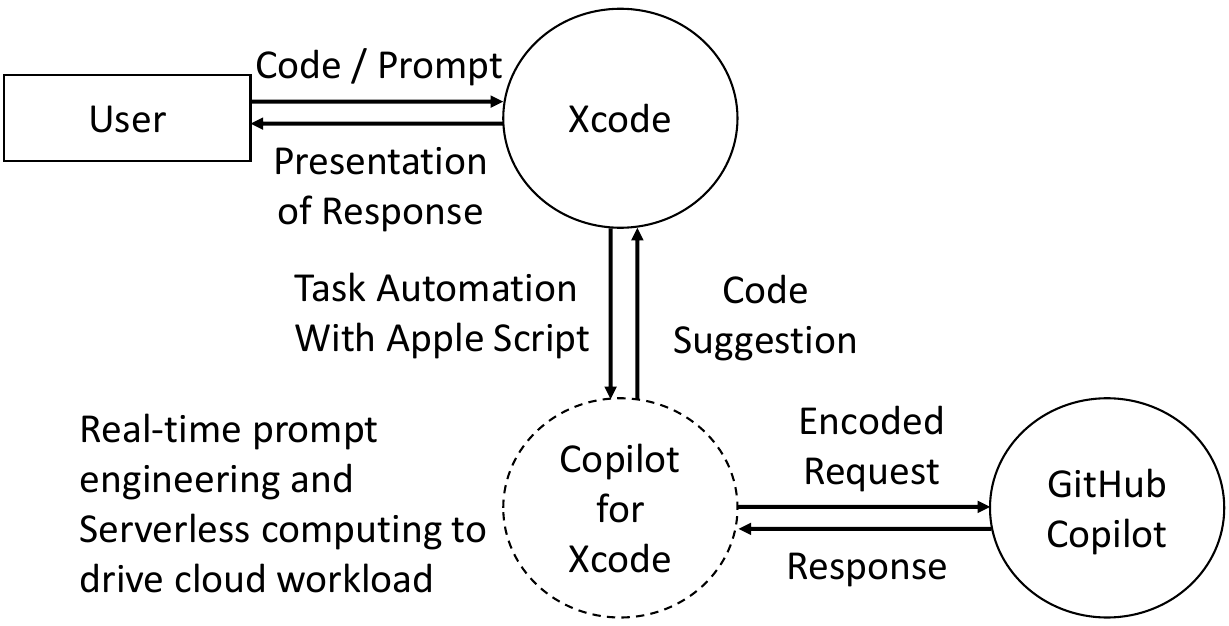}
\caption{An overview of the AI-assisted programming application, {\it Copilot for Xcode} modeled as an intermediary software entity to connect user requests (e.g., prompt tokens) with cloud-based large language models. }\label{fig:overview}
\end{figure*}

\section{Introduction}
The field of natural language processing (NLP) has witnessed remarkable achievements through the use of large language models (LLMs). These models exhibit remarkable skills in understanding and generating natural languages. Additionally, they employ feedback mechanisms, such as rewards or penalties, to improve their comprehension and fine-tune their future performance \cite{christiano2017deep,ouyang2022training}. The application of LLMs to AI-assisted programming has recently attracted considerable attention
\cite{rajamani2022ai,wong2023natural}, as it offers the possibility to embed advanced conversational agents in software development \cite{li2022automating,chen2021evaluating}.  In fact, the emergence of LLM-driven tools like ChatGPT (Chat Generative Pre-Trained Transformer), Github Copilot, DeepMind's AlphaCode resonates with the visionary ideas presented in Edsger W. Dijkstra's seminal paper in \cite{dijkstra}, illustrating the transformative potential of computers in facilitating a seamless integration of code and creativity. By surpassing the boundaries of debugging, AI-assisted programming tools can embrace the harmonious combination of program composition and elegant design  \cite{dijkstra}. 

Since the work in \cite{dijkstra}, one of the earliest AI-assisted programming tool is the {\it MIT programmer's apprentice}, which aimed to simulate a knowledgeable junior programmer and utilized natural language processing to acquire understanding of programming patterns, clichés, and interactions \cite{MITapprentice82,MITapprentice88}. The ``MIT programmer's apprentice" played a pioneering role in introducing revolutionary concepts such as code generation (e.g., see \cite{MITcodegeneration}) and an early form of "prompt engineering" (e.g., see \cite{MITprompt}). These advancements were driven by the recognition of computer programming as a systematic process of abstraction and simplification \cite{dijkstra,MITsimplify}.

AI-assisted programming improves software productivity by automating tasks, detects errors, enhances code quality, promotes usability, improves reliability and accelerates the overall software development cycles \cite{wong2023natural}. Rather than replacing human programmers, these tools empower them to unleash their creative potential. By automating repetitive and mundane tasks, AI-assisted programming frees up valuable time and mental energy for human programmers to focus on innovative problem-solving and designing elegant solutions with the help of predictive analysis. Furthermore, by incorporating natural language processing capabilities (i.e., via prompt engineering), these tools enable human programmers to interact with software systems in a more intuitive and human-like manner, thus streamlining the software development process~\cite{dijkstra1972humble}. 

Cloud-based tools that leverage LLMs such as Codeium~\cite{codeium}, GitHub Copilot~\cite{friedman2021introducing}, OpenAI ChatGPT~\cite{openai_2023}, and Amazon CodeWhisperer~\cite{amazon}, enable users to access their cloud-based LLM services and online resources through dedicated application programming interface (API) in an on-demand access. The pricing models for these tools vary depending on the complexity of the tool and the target audience. Some pricing models include enterprise pricing, subscription-based pricing, usage-based pricing, freemium (i.e., the tool is available for free, but additional premium features require payment), pay-per-use pricing or entirely free of charge. In fact, these pricing models and LLM-based services can be incorporated into existing systems like a local integrated development environment (IDE) or implemented via a Software-as-a-Service (SaaS) web interface, acting as a virtual service entity to meeting objectives and saving costs for the human programmer
\cite{zheng2015bid,zheng2016viability}. It is expected that the expanding reach and high demand usage of these LLM-based tools reflect the growing need for advanced NLP capabilities in software development. This trend aligns with Dijkstra's visionary ideas as discussed in \cite{dijkstra1972humble,dijkstra}.

This paper presents {\it Copilot for Xcode}, an AI-assisted programming tool that was open-sourced on December 7, 2022, one week after OpenAI launched its ChatGPT on November 30, 2022.\footnote{Apple's recently-issued patent in \cite{applepatent} dated June 27, 2023 suggests that they are actively exploring the integration of machine learning models into their software development system, specifically within Xcode, rather than relying solely on existing solution.} Acting as an intermediary entity, as shown in Figure~\ref{fig:overview}, it seamlessly integrates cloud-based large language model services with local IDEs like Xcode. This integration benefits software developers in the Apple ecosystem by streamlining AI-assisted programming service delivery and enhancing the accessibility of a myriad of cloud-based LLM applications. {\it Copilot for Xcode} enables real-time prompt engineering and efficient interaction between the human programmer and the large language models, offering the potential to integrate serverless computing capabilities with natural language processing in the cloud. The source code of {\it Copilot for Xcode} can be publicly accessed at \url{https://github.com/intitni/CopilotForXcode}.

% \begin{itemize}
%     \item We present an AI-assisted programming application, Copilot for Xcode, that supports the real-time prompt engineering and serverless computing which can be publicly accessed on \url{https://github.com/intitni/CopilotForXcode}.
%     \item We present comprehensive case studies that showcase its practical application to real-world problems.
% \end

\begin{figure*}[!tb]
\centering
\includegraphics[width=0.8\linewidth]{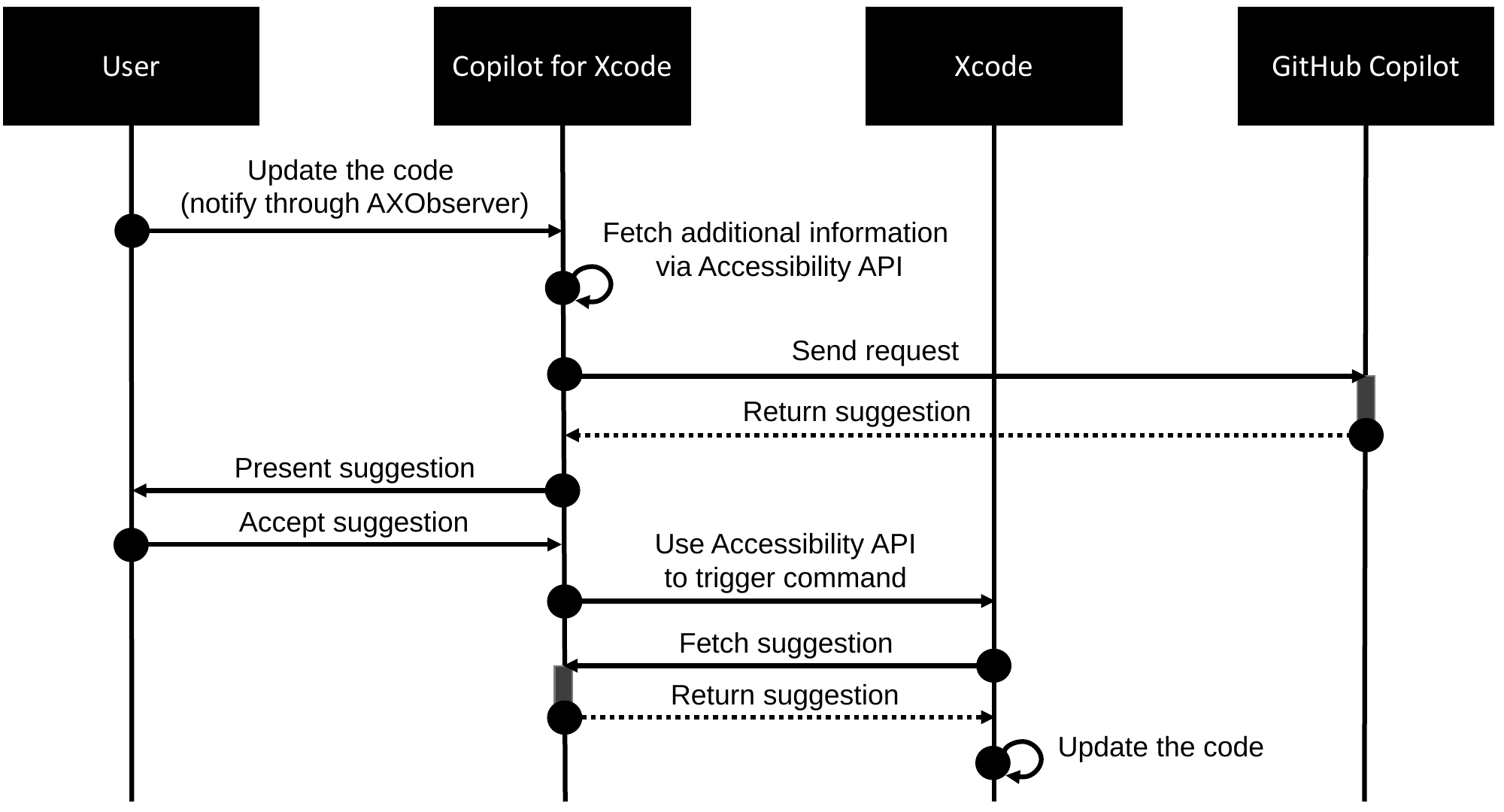}
\caption{The sequence diagram illustrates the functionality of {\it Copilot for Xcode}, enabling real-time suggestions through integration with GitHub Copilot. When a user initiates a code update, {\it Copilot for Xcode} receives a notification and subsequently sends a request to the GitHub Copilot API. Upon receiving the suggestions from GitHub Copilot, the user has the option to adopt the recommendations and directly apply the changes within Xcode.}\label{fig:copilot_for_xcode_fetch_suggestions}
\end{figure*}

\section{Related Works}
\subsection{Language Models for Big Code Analysis}
LLMs have surfaced as a promising approach to tackle challenges in computer programming, leveraging the software naturalness hypothesis \cite{hindle2012naturalness}. This hypothesis posits that programming languages can be understood and manipulated in a similar fashion to how natural language processing techniques handle human languages. Since the introduction of the transformer architecture in 2017 \cite{vaswani2017attention}, LLMs trained on large-scale datasets of programs have shown significant benefits in code-related tasks by effectively learning programming language patterns and structures, which are collectively part of Big Code analysis \cite{vechev2016programming}. Recent LLMs such as T5~\cite{raffel2020exploring}, BERT~\cite{devlin2018bert}, GPT-4~\cite{gpt42023} and Palm 2~\cite{anil2023palm} have demonstrated impressive capabilities in understanding and generating human-like text, opening up new possibilities for enhancing software engineers' development experiences. These models undergo a two-step process involving pre-training and fine-tuning. Following these steps, prompt engineering can be applied to further optimize the model's performance. As an integral part of AI-assisted programming, AI-based predictive analysis~\cite{ji2020amazing} can anticipate potential issues in a software development life cycle. For example, it can proactively identify and flag critical incidents~\cite{surameery2023use} before they manifest~\cite{talamadupula2021applied}.

\subsection{AI-assisted Programming}
AI-assisted programming is the incorporation of machine learning techniques and tools into the software development process~\cite{mozannar2022reading} to improve computer programming tasks. This concept shares similarities with pair programming~\cite{bird2022taking,imai2022github}, whereby two human programmers collaborate to develop software by alternating between writing code (driver) and reviewing (observer) in a continuous switch. AI-assisted programming essentially replaces one of the two human programmers with an AI assistant, akin to the aforementioned ``MIT programmer's apprentice" \cite{MITapprentice82,MITapprentice88}. The AI assistant automates tasks that can be broadly classified into two categories: generation and understanding. Generation tasks encompass activities such as code generation \cite{waldinger1969prow,manna1971toward}, code completion \cite{robbes2008program,bruch2009learning}, code translation \cite{acharya2007mining,allamanis2014naturalize}, code refinement \cite{saha2017elixir}, and code summarization \cite{sridhara2010towards,sridhara2011generating}. On the other hand, understanding tasks encompass activities like defect detection \cite{charniak1996statistical} and clone detection \cite{kontogiannis1996pattern}. Improving the quality of large language models for these tasks focus on enhancing pre-training schemes \cite{li2022automating}, expanding training corpora \cite{husain2019codesearchnet}, and employing improved evaluation metrics \cite{chen2021evaluating}. The GitHub Copilot~\cite{friedman2021introducing} is an example of an AI-powered programming tool that utilizes OpenAI Codex, which is based on GPT-3 LLM that has been trained on a vast amount of source code from the GitHub repository, totaling over 159GB \cite{openai_2023}. For further details on the latest advancements in AI-assisted programming, please see \cite{wong2023natural}.

\begin{figure*}[!tb]
\centering
\includegraphics[width=\linewidth]{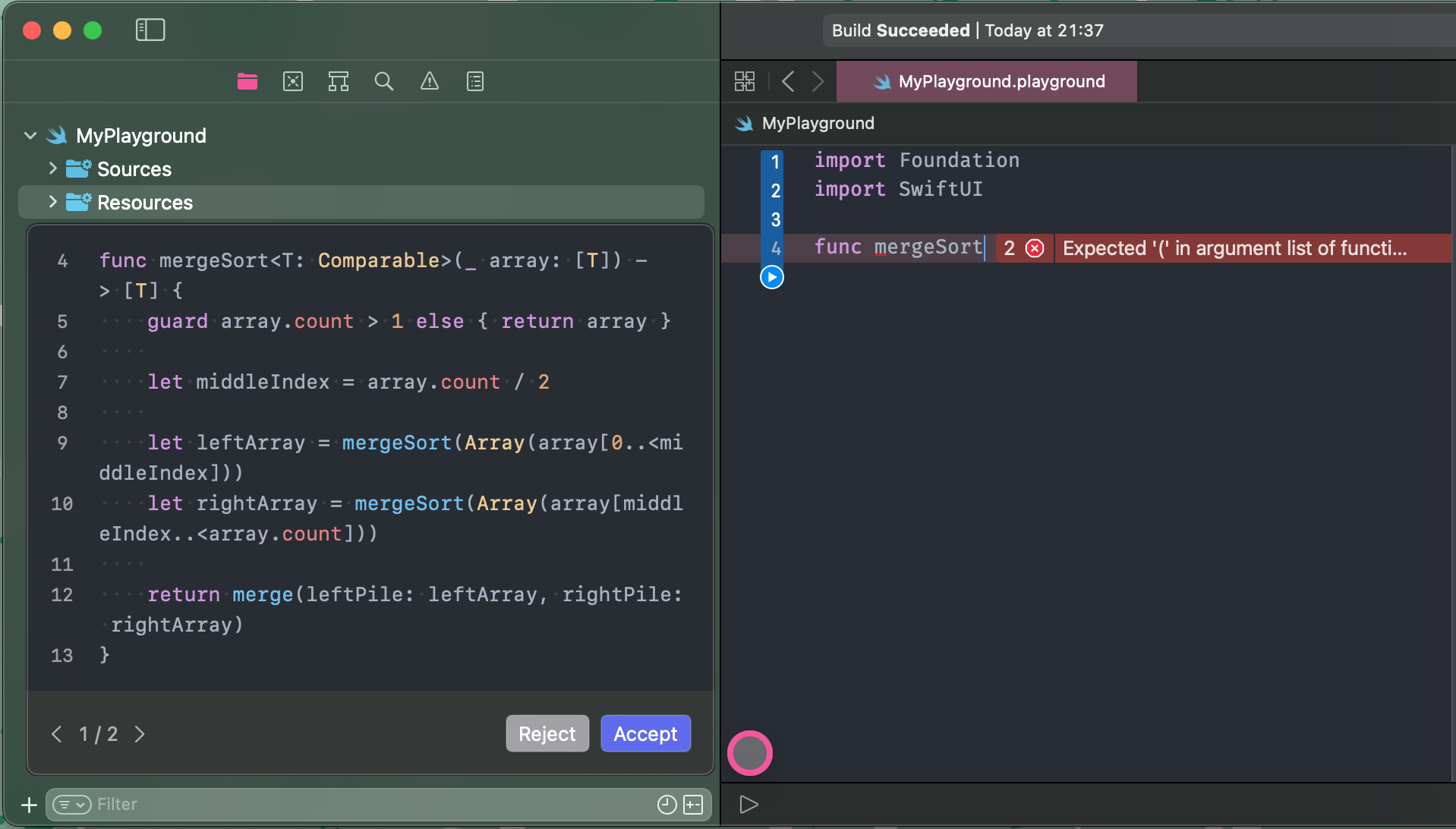}
\caption{The user interface of {\it Copilot for Xcode} demonstrates the code suggestion capability, specifically showcasing the integration of GitHub Copilot for real-time suggestions related to the merge sort algorithm. In the accompanying Figure, the right-hand side displays the open source code editor within Xcode, focused on an interactive Swift playground. On the left-hand side, the cloud-based services deliver code suggestion responses.}
\label{fig:copilot_for_xcode_screenshot_suggestion}
\end{figure*}

\section{Copilot for Xcode}
\subsection{Xcode and its Offline Functionalities}
Xcode~\cite{xcode} is an IDE created by Apple for developing software applications for the Apple ecosystem such as macOS and iOS. It provides a comprehensive set of tools, including editors, compilers, debugging tools, and interface builders, to help software developers create and maintain their applications. Xcode includes a source code editor with features like syntax highlighting, code completion, and refactoring capabilities. It supports multiple programming languages, including Swift, Objective-C, C, and C++, allowing developers to write code for a variety of Apple platforms. In addition to the code editor, Xcode offers a wide range of tools to assist in app development, such as an Interface Builder for designing user interfaces visually, a graphical debugger for finding and fixing issues in code, and various performance analysis instruments. It also integrates with other developer tools, such as the iOS Simulator, which allows software developers to test their apps on virtual devices, and Instruments, a powerful profiling tool for measuring and optimizing app performance. Despite the extensive functionalities of Xcode, it has some limitations. For example, certain features depend on an offline rule-based system and may necessitate batch updates from Apple. Consequently, these services may not remain up-to-date consistently. 

\subsection{The Copilot for Xcode Framework}
The main limitation in Xcode is the sandboxing mechanism, which restricts plugin access to specific resources and prevents the launching of other programs. We address ways to overcome this limitation in order to enable the functionality of GitHub Copilot in Xcode. In particular, GitHub Copilot requires an additional program provided by GitHub to be executed alongside the plugin. Let us take the real-time suggestion feature as an example: the application first needs to bypass the sandbox in order to run the GitHub Copilot language server. This is accomplished by establishing communication between the Xcode source editor extension and a non-sandboxed XPC Service, which acts as a cross-process call service that facilitates the communication between the extension and the GitHub Copilot server. The server then presents suggestions in a user interface (UI) that is not managed by Xcode. To assemble a request for the language server, the application must gather sufficient information, but Xcode only provides the source code and file type. To obtain additional information without relying on Xcode, the application leverages the {\tt Accessibility API} from the software development kit. This particular API exposes information about each object within the application. Furthermore, to enable in-place code editing, the application executes extension commands programmatically. This is accomplished by utilizing the {\tt Accessibility API} to interact with the menu bar items. These implementations thus allow Apple software developers to leverage GitHub Copilot~\cite{friedman2021introducing} and Codeium~\cite{codeium} for code suggestions, while utilizing ChatGPT~\cite{openai_2023} for code explanations, generation and natural language-based code modifications. The technical interaction of integrating Copilot with Xcode are depicted in Figure~\ref{fig:copilot_for_xcode_fetch_suggestions}. 

In addition, it facilitates the integration of an external chat panel that can access and read the user's code. This chat panel serves as a connection point to leverage LLMs for functionalities such as code explanation and mutation using natural language. The chat panel can also be extended with plugins to offer additional features, including support for answering questions with real-time information from search engines. Some latest cloud-based LLM provide direct access through their official APIs for direct integration. In particular, {\it Copilot for Xcode} leverages the LangChain~\cite{Langchain} framework, which facilitates the creation of customized LLMs tailored to specific use cases. This framework significantly enhances the prompt engineering process~\cite{wu2022promptchainer,poldrack2023ai}, allowing for the design of more effective prompts that can be utilized with the LLMs. This integration and framework combination optimize the functionality and usability of the LLMs, providing users with enhanced capabilities and improved prompt customization.

\begin{figure*}[!tb]
\centering
\includegraphics[width=\linewidth]{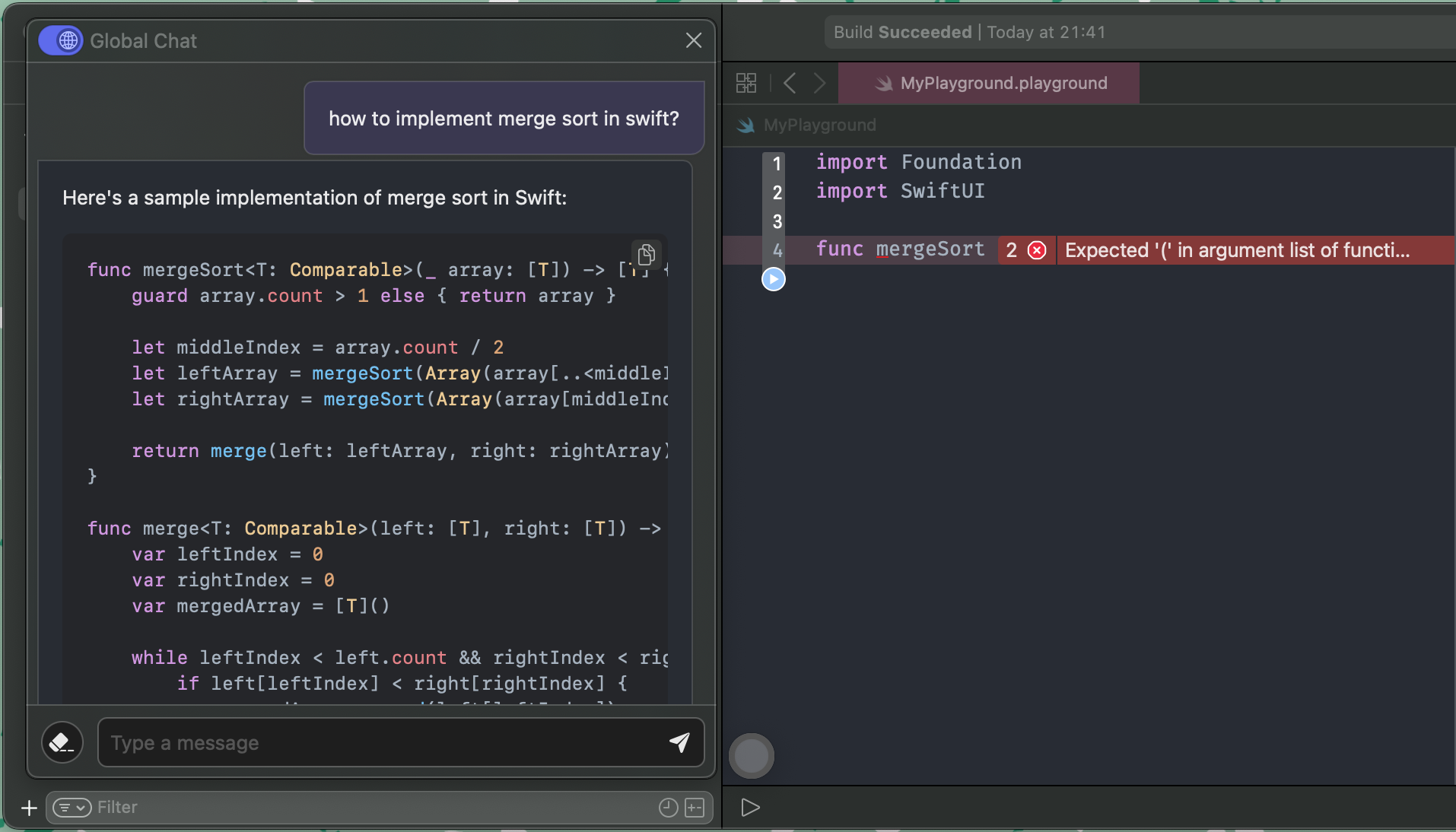}
\caption{The user interface of {\it Copilot for Xcode} shows its Chat and prompt-to-code features, which enable code generation for the merge sort algorithm. These features are connected to ChatGPT, allowing for online prompt engineering and code generation within Xcode. The Figure illustrates the source code editor in Xcode on the right-hand side, while the chat conversational panel is displayed on the left-hand side.} \label{fig:copilot_for_xcode_screenshot_chat}
\end{figure*}

\subsection{Code Suggestion}\label{sec:acs}
The code suggestion function offers a viable option for code completion and generation under diverse usage scenarios. Code completion, commonly referred to as auto-completion~\cite{wong2023natural}, is an invaluable feature in software development that assists in completing unfinished code segments. On the other hand, code generation involves the automatic generation of source code from natural language input~\cite{li2022competition}, guided by user-defined constraints. This capability strengthens the efficiency on the development process by automating the creation of code based on linguistic specifications provided by the user. 

In {\it Copilot for Xcode}, we offer real-time code suggestions that dynamically update whenever users modify their code. This capability, depicted in Figure~\ref{fig:copilot_for_xcode_screenshot_suggestion}, is powered by the integration of GitHub Copilot and Codeium, ensuring that the suggestions are specifically tailored to the files currently open in the workspace, thus enhancing productivity and accuracy while leveraging the capabilities of the code suggestion function. The feature offers two presentation modes for displaying suggestions, which includes two distinct modes. In the ``Nearby Text Cursor'' mode, suggestions are presented based on the current position of the text cursor. On the other hand, the ``Floating Widget'' mode displays suggestions in close proximity to the circular widget. When the user updates their code, the integrated application retrieves and integrates relevant suggestions for display within Xcode.

The software development experience is further enhanced by a range of pre-defined commands offered by {\it Copilot for Xcode}. The first useful command is {\tt Get Suggestions}, which retrieves customized suggestions based on the current cursor position in the edited file on Xcode. In cases where multiple suggestions are available, users can conveniently navigate through them using the {\tt Next Suggestion} and {\tt Previous Suggestion} commands to choose the code suggestions based on their preferences. When incorporating suggested code, the {\tt Accept Suggestion} command comes in handy to immediately select the code suggestion, while the {\tt Reject Suggestion} command allows users to remove unnecessary suggestions along with their associated comments. Furthermore, there are two commands specifically designed for the usage of {\it Copilot for Xcode}. The {\tt Real-time Suggestions} command, which can only be called by the {\it Copilot for Xcode} automatically, provides real-time suggestions after a successful retrieval so that the code suggestion can be presented in Xcode, while the {\tt Prefetch Suggestions}, which can also be called by the {\it Copilot for Xcode}, command proactively fetches real-time suggestions in the background, thus improving the overall responsiveness.

\subsection{Chat and Prompt-to-Code for Code Generation}
{\it Copilot for Xcode} provides its chat and prompt-to-Code features for code generation, as depicted in Figure~\ref{fig:copilot_for_xcode_screenshot_chat}. This functionality focuses on generating code from text inputs, enabling text-to-code generation within the IDE. By incorporating these advanced code generation capabilities,{\it Copilot for Xcode} enhances coding workflows, making them more efficient and intuitive. The chat function of our application, also powered by ChatGPT, complements these code generation features and offers additional enhancements for a interactive coding experience. Users can leverage specific features customized to their programming needs, such as extracting selected code in the active editor for reference and discussion of specific code snippets. Access to the relative path of the file being worked on facilitates easy navigation within the codebase. The chat or prompt-to-code functions also assist in capturing error and warning labels in the active editor, enabling swift issue resolution. Users can also obtain information about the text cursor location, facilitating precise discussions and context-aware conversations. These combined features empower users to engage in productive coding discussions and streamline their coding process, harnessing the capabilities of our AI-powered application.

The prompt-to-Code function offers a range of capabilities for code modification and creation using natural language. It is particularly beneficial when there is a need to update a specific section of code. This feature provides various use cases, such as enhancing code readability, rectifying code bugs, including documentation within the code, dividing extensive functions into smaller, manageable ones, generating code based on specific templates using custom commands, refining grammar and spelling errors in documentation, and facilitating the translation of localizable strings files. With "Prompt to Code," users can refactor existing code or write new code by harnessing the power of natural language.

\section{Evaluation}
We describe three case studies that illustrate the power of {\it Copilot for Xcode} in tackling real-world programming challenges through AI-assisted programming. The case studies presented here are based on real-world programming assignments given to undergraduate students. Furthermore, the case studies also highlight the significance of prompt engineering for code suggestion query and making ``small" decisions for program composition and design in {\it Copilot for Xcode}.

\subsection{Case Study: HCF of Two Numbers}
The first case study considers computing the Highest Common Factor (HCF) or the Greatest Common Divisor (GCD) of two natural numbers \cite{dijkstraeuclid}. The HCF/GCD represents the largest positive integer that divides both numbers, leaving no remainder. Many existing approaches can be used to solve this problem, including subtraction, brute-force, and binary algorithm. One of the oldest algorithms to compute the HCF for two given natural numbers is the Euclidean algorithm. We tasked students to implement the Euclidean algorithm using the Swift programming language based on the observation that if $r$ is the remainder when $a$ is divided by $b$, then the HCF of $a$ and $b$ is equivalent to the HCF of $b$ and $r$. Figure 5 depicts the brute-force approach for the HCF of two natural numbers, while Figure 6 provides a correct depiction of the HCF calculation using the Euclidean algorithm.

\begin{table}[!h]
  \centering
  \begin{tabular}{ @{\hskip3pt}l@{\hskip3pt} @{\hskip3pt}l@{\hskip3pt}} %p{2cm}p{5.8cm}  }
    % \hline
    \textbf{Prompt:} & {\tt HCF of Two Numbers}\\
    \textbf{Result:} & \\
      & \begin{minipage}{0.39\textwidth}
                        \vspace{-0.3cm}
                        \includegraphics[width=\linewidth]{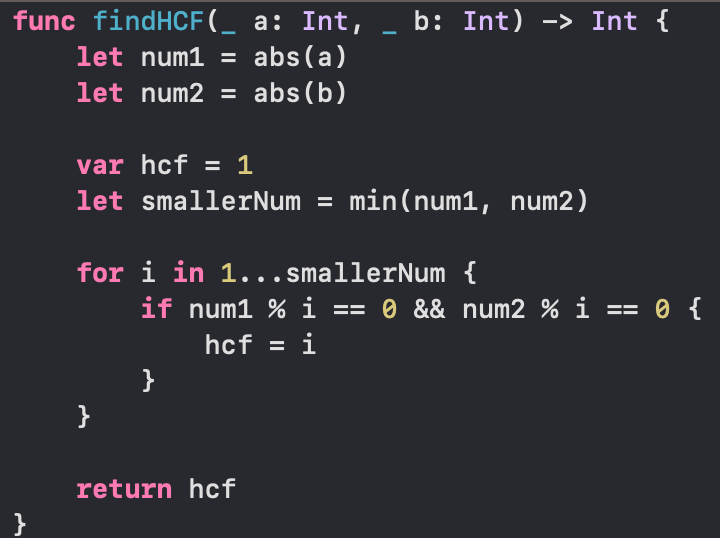}

                        \end{minipage}\\
  \end{tabular}
  \captionsetup{labelformat=empty}
                       \caption{Figure 5: Code suggestion of HCF of two natural numbers using a brute-force algorithm without specific instructions in the prompt.}
 \label{fig:case0}
\end{table}

\begin{table}[!h]
  \centering
  \begin{tabular}{ @{\hskip3pt}l@{\hskip3pt} @{\hskip3pt}l@{\hskip3pt}} %p{2cm}p{5.8cm}  }
    % \hline
    \textbf{Prompt:} & {\tt HCF of Two Numbers by Euclidean} \\
    & {\tt Algorithm} \\
    \textbf{Result:} & \\
      & \begin{minipage}{0.39\textwidth}
                        \vspace{-0.3cm}
                        \includegraphics[width=\linewidth]{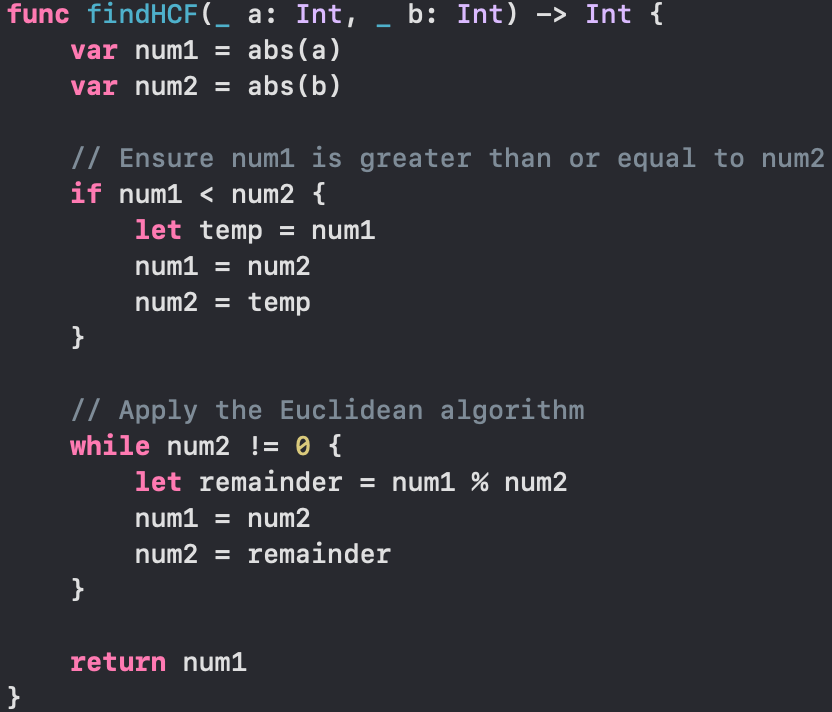}
                        \end{minipage}\\
  \end{tabular}
  \captionsetup{labelformat=empty}
                       \caption{Figure 6: Code suggestion of HCF of two natural numbers with specific instructions on using the Euclidean algorithm.}
 \label{fig:case1}
\end{table}
\subsection{Case Study: LCM of Two Numbers}
The Least Common Multiple (LCM) of two natural numbers $a,b$ refers to the smallest positive integer that is divisible by both numbers. Notably, this LCM is dual to the HCF/GCD \cite{dijkstraeuclid}.\footnote{The product of two natural numbers is equal to the product of their respective least common multiple and greatest common denominator. This principle can be demonstrated using the Fundamental Theorem of Arithmetic in number theory or through an algorithmic method outlined in \cite{dijkstraeuclid}. From our observations, the ChatGPT Codex (GPT-3) was able to understand this concept (even able to demonstrate a plausible proof with prompt engineering), although it faced difficulties in extending the duality to encompass more than two natural numbers.} Typically, the LCM algorithm makes use of the HCF algorithm. However, in this assignment, a unique requirement is to develop an LCM algorithm that does not rely on the HCF algorithm. By default, the code suggestion in {\it Copilot for Xcode} assumes that developers have already implemented the HCF function without any additional instructions in the prompt. The tool thus generates a result that utilizes the HCF implementation as a helper function for the LCM as shown in Figure 7. However, to comply with the specific prompt of not relying on the HCF, Figure 8 presents a correct answer for the LCM calculation without requiring to compute the HCF.
\begin{table}[!h]
  \centering
  \begin{tabular}{ @{\hskip3pt}l@{\hskip3pt} @{\hskip3pt}l@{\hskip3pt}} %p{2cm}p{5.8cm}  }
    % \hline
    \textbf{Prompt:} & {\tt LCM of Two Numbers} \\
    \textbf{Result:} & \\
      & \begin{minipage}{0.39\textwidth}
                        \vspace{-0.3cm}
                        \includegraphics[width=\linewidth]{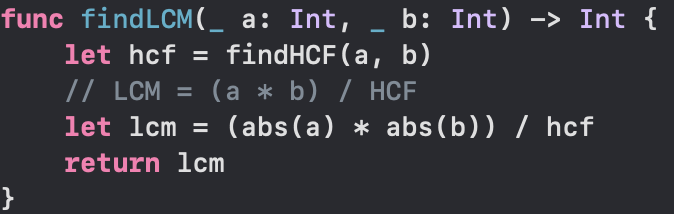}

                        \end{minipage}\\
  \end{tabular}
  \captionsetup{labelformat=empty}
  \caption{Figure 7: Code suggestion of LCM of two natural numbers using the HCF without specific instructions.}
 \label{fig:case2}
\end{table}

\begin{table}[!h]
  \centering
  \begin{tabular}{ @{\hskip3pt}l@{\hskip3pt} @{\hskip3pt}l@{\hskip3pt}} %p{2cm}p{5.8cm}  }
    % \hline
    \textbf{Prompt:} & {\tt LCM of Two Numbers without}\\
    & {\tt Using the HCF}\\
    \textbf{Result:} & \\
      & \begin{minipage}{0.39\textwidth}
                        \vspace{-0.3cm}
                        \includegraphics[width=\linewidth]{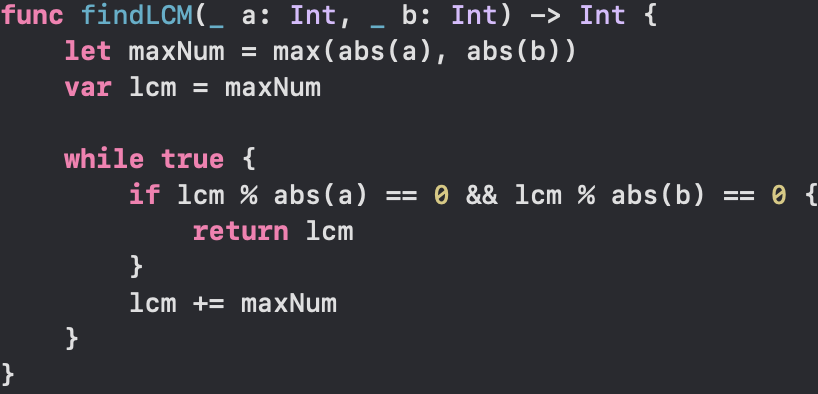}

                        \end{minipage}\\
  \end{tabular}
  \captionsetup{labelformat=empty}
  \caption{Figure 8: Code suggestion of LCM of two natural numbers without using HCF algorithm as a helper function.}
 \label{fig:case3}
\end{table}

\subsection{Case Study: Navigating App on iOS}
In the example below, we delve into a code generation scenario that effectively highlights the fundamental concepts of SwiftUI in iOS app development, specifically focusing on view navigation. This scenario is visually represented in Figure 9. The iOS app, built with SwiftUI, comprises two distinct views: HomeView and DetailView, as depicted in Figure 10. View refer to a crucial component that constructs the UI and plays a pivotal role in displaying and handling the visual elements that users observe and engage with on the screen. To effectively manage views, software developers are required to arrange them in a hierarchical structure and personalize each view by configuring different properties. 

\begin{table}[!h]
  \centering
  \begin{tabular}{ @{\hskip3pt}l@{\hskip3pt} @{\hskip3pt}l@{\hskip3pt}} 
    \textbf{Prompt:} & {\tt Create a navigating views app} \\
    & {\tt with SwiftUI}\\
    \textbf{Result:} & \\
      & \begin{minipage}{0.39\textwidth}
                        \vspace{-0.3cm}
                        \includegraphics[width=\linewidth]{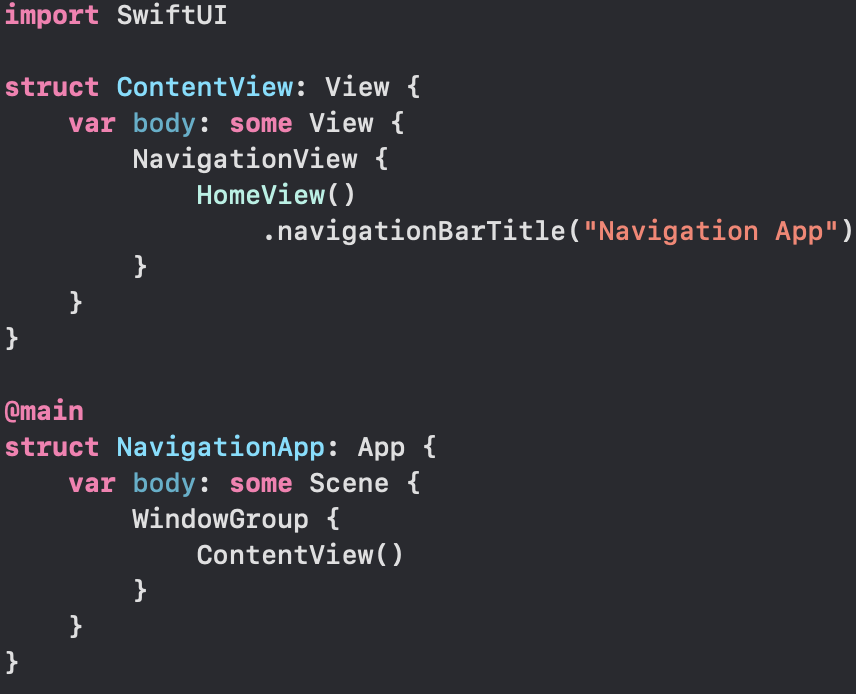}

                        \end{minipage}\\
  \end{tabular}
  \captionsetup{labelformat=empty}
  \caption{Figure 9: This SwiftUI-based app consists of two screens: a home screen and a detail screen. The ContentView sets up a navigation with the entry of the app and provides a navigation bar title.}
 \label{fig:case4_1}
\end{table}

\begin{table}[!h]
  \centering
  \begin{tabular}{ @{\hskip3pt}l@{\hskip3pt} @{\hskip3pt}l@{\hskip3pt}} 
    \textbf{Prompt:} & {\tt Create the HomeView and } \\
    & {\tt DetailsView with SwiftUI} \\
    \textbf{Result:} & \\
      & \begin{minipage}{0.39\textwidth}
                        \vspace{-0.3cm}
                        \includegraphics[width=\linewidth]{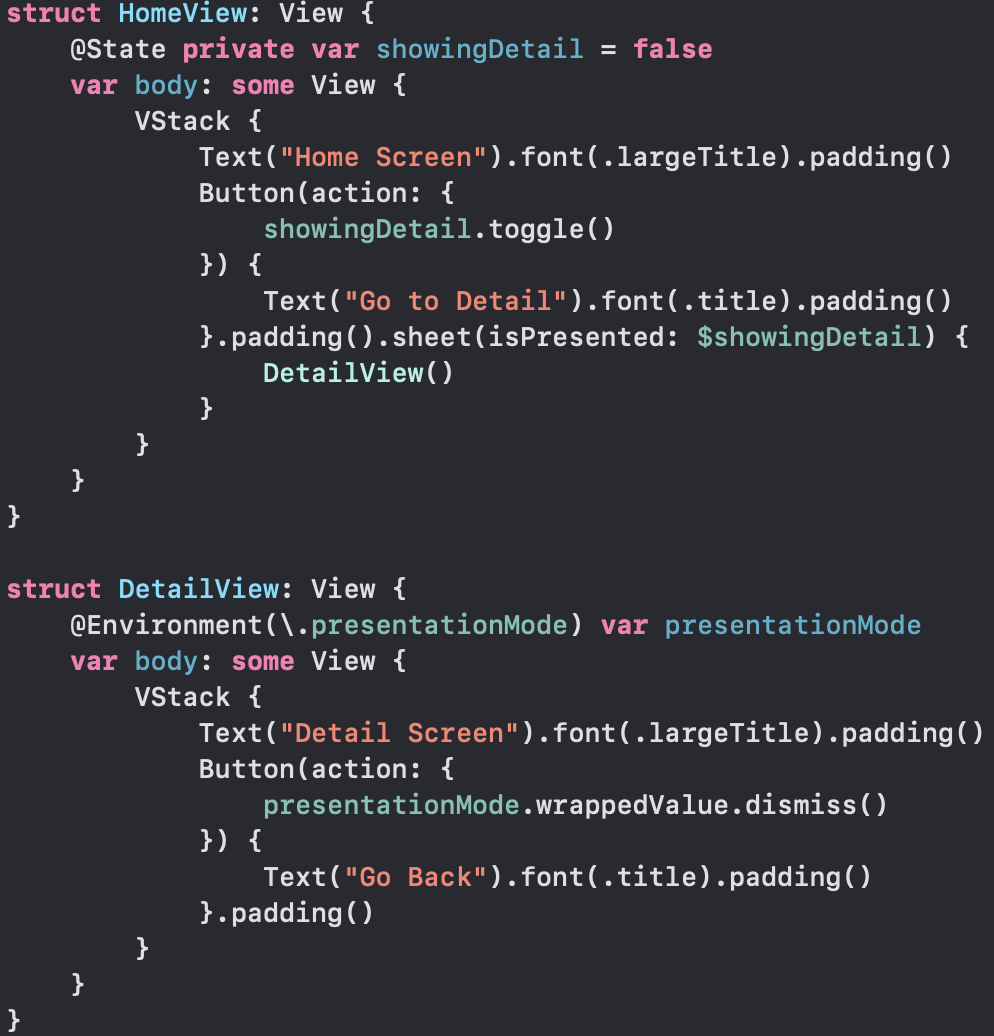}

                        \end{minipage}\\
  \end{tabular}
  \captionsetup{labelformat=empty}
  \caption{Figure 10: The home and detail screen for a navigating app. When a button on the home screen is tapped, the app navigates to the detail screen. Additionally, a back button on the detail screen allows the user to navigate back to the home screen.}
 \label{fig:case4_2}
\end{table}

\section{Conclusion}
This paper introduced {\it Copilot for Xcode} that integrates cloud-based large language model services (Github Copilot and OpenAI's GPT) with Apple's integrated development environment, Xcode, for AI-assisted programming. We also discussed the efficacy of prompt engineering and possible strategies for AI-assisted programming using simple case studies to illustrate the practical application of this tool to program composition and design. When designing a program, making small decisions often involves breaking down complex tasks into smaller components manageable by the large language model. By carefully constructing prompts, programmers can influence the generation of code and steer the langage model's understanding towards the desired outcome.

As a software prototype, {\it Copilot for Xcode} has some limitations to consider during practical usage. For example, to bypass the sandboxing restrictions, it employs unconventional methods to retrieve information like file and project/workspace paths. As such, it is important to be aware that this might not always function seamlessly in future versions of Xcode. Also, the current code suggestions are presented as C-style comments in comment mode, which can inadvertently disrupt a user's code if they are working on a format, e.g., JSON file, where such comments are not applicable.

By combining the capabilities of large language models and integrated tools for prompt engineering, {\it Copilot for Xcode} enhances and streamlines the software development process within Apple's Xcode. The integration of {\it Copilot for Xcode} with other cloud-based services like Xcode Cloud can also improve the overall productivity and efficiency in software development, which is especially important to continuous integration (CI) and continuous delivery (CD) in the software development pipeline. As AI-assisted programming tools like Copilot get incorporated into more IDEs, it brings us closer to the realization of Dijkstra's vision in \cite{dijkstra}, fostering a symbiotic relationship between human programmers and AI-powered tools to achieve more efficient and reliable software development.

\bibliographystyle{named}
\bibliography{ijcai23}

\begin{thebibliography}{}

\bibitem[\protect\citeauthoryear{Acharya \bgroup \em et al.\egroup
  }{2007}]{acharya2007mining}
Mithun Acharya, Tao Xie, Jian Pei, and Jun Xu.
\newblock Mining api patterns as partial orders from source code: From usage
  scenarios to specifications.
\newblock In {\em 6th Joint Meeting of The European Software Engineering
  Conference and The ACM SIGSOFT Symposium on The Foundations of Software
  Engineering}, pages 25--34, 2007.

\bibitem[\protect\citeauthoryear{Allamanis \bgroup \em et al.\egroup
  }{2014}]{allamanis2014naturalize}
Miltiadis Allamanis, Earl~T. Barr, Christian Bird, and Charles Sutton.
\newblock Learning natural coding conventions.
\newblock In {\em 22nd ACM SIGSOFT International Symposium on Foundations of
  Software Engineering}, page 281–293. Association for Computing Machinery,
  2014.

\bibitem[\protect\citeauthoryear{Amazon}{2022}]{amazon}
CodeWhisperer Amazon.
\newblock {AI} code generator - amazon codewhisperer.
\newblock \url{https://aws.amazon.com/codewhisperer}, 2022.
\newblock Accessed on June 1, 2023.

\bibitem[\protect\citeauthoryear{Anil \bgroup \em et al.\egroup
  }{2023}]{anil2023palm}
Rohan Anil, Andrew~M Dai, Orhan Firat, Melvin Johnson, Dmitry Lepikhin,
  Alexandre Passos, Siamak Shakeri, Emanuel Taropa, Paige Bailey, Zhifeng Chen,
  et~al.
\newblock Palm 2 technical report.
\newblock {\em arXiv preprint arXiv:2305.10403}, 2023.

\bibitem[\protect\citeauthoryear{Apple}{2003}]{xcode}
Xcode Apple.
\newblock Xcode 15 - apple developer.
\newblock \url{https://developer.apple.com/xcode/}, 2003.
\newblock Accessed on June 1, 2023.

\bibitem[\protect\citeauthoryear{Bird \bgroup \em et al.\egroup
  }{2022}]{bird2022taking}
Christian Bird, Denae Ford, Thomas Zimmermann, Nicole Forsgren, Eirini
  Kalliamvakou, Travis Lowdermilk, and Idan Gazit.
\newblock Taking flight with copilot: Early insights and opportunities of
  {AI}-powered pair-programming tools.
\newblock {\em Queue}, 20(6):35--57, 2022.

\bibitem[\protect\citeauthoryear{Bruch \bgroup \em et al.\egroup
  }{2009}]{bruch2009learning}
Marcel Bruch, Martin Monperrus, and Mira Mezini.
\newblock Learning from examples to improve code completion systems.
\newblock In {\em 7th Joint Meeting of The European Software Engineering
  Conference and The ACM SIGSOFT Symposium on The Foundations of Software
  Engineering}, pages 213--222, 2009.

\bibitem[\protect\citeauthoryear{Charniak}{1996}]{charniak1996statistical}
Eugene Charniak.
\newblock {\em Statistical Language Learning}.
\newblock MIT press, 1996.

\bibitem[\protect\citeauthoryear{Chase}{2022}]{Langchain}
Harrison Chase.
\newblock Langchain, 2022.

\bibitem[\protect\citeauthoryear{Chen \bgroup \em et al.\egroup
  }{2021}]{chen2021evaluating}
Mark Chen, Jerry Tworek, Heewoo Jun, Qiming Yuan, Henrique Ponde de~Oliveira
  Pinto, Jared Kaplan, Harri Edwards, Yuri Burda, Nicholas Joseph, Greg
  Brockman, et~al.
\newblock Evaluating large language models trained on code.
\newblock {\em arXiv preprint arXiv:2107.03374}, 2021.

\bibitem[\protect\citeauthoryear{Christiano \bgroup \em et al.\egroup
  }{2017}]{christiano2017deep}
Paul~F Christiano, Jan Leike, Tom Brown, Miljan Martic, Shane Legg, and Dario
  Amodei.
\newblock Deep reinforcement learning from human preferences.
\newblock {\em Advances in neural information processing systems}, 2017.

\bibitem[\protect\citeauthoryear{Codeium}{2023}]{codeium}
Exafunction Codeium.
\newblock Codeium - free {AI} code completion \& chat.
\newblock \url{https://codeium.com/}, 2023.
\newblock Accessed on June 1, 2023.

\bibitem[\protect\citeauthoryear{Devlin \bgroup \em et al.\egroup
  }{2018}]{devlin2018bert}
Jacob Devlin, Ming-Wei Chang, Kenton Lee, and Kristina Toutanova.
\newblock Bert: Pre-training of deep bidirectional transformers for language
  understanding.
\newblock {\em arXiv preprint arXiv:1810.04805}, 2018.

\bibitem[\protect\citeauthoryear{Dijkstra}{1972}]{dijkstra1972humble}
Edsger~W Dijkstra.
\newblock The humble programmer.
\newblock {\em Communications of the ACM}, 15(10):859--866, 1972.

\bibitem[\protect\citeauthoryear{Dijkstra}{2007}]{dijkstraeuclid}
Edsger~Wybe Dijkstra.
\newblock Defining the greatest common divisor.
\newblock {\em E. W. Dijkstra Archive (EWD 1257)}, 2007.

\bibitem[\protect\citeauthoryear{Dijkstra}{transcribed 2007}]{dijkstra}
Edsger~Wybe Dijkstra.
\newblock A preliminary investigation into computer assisted programming.
\newblock {\em E. W. Dijkstra Archive (EWD 237)}, (transcribed) 2007.

\bibitem[\protect\citeauthoryear{Friedman}{2021}]{friedman2021introducing}
Nat Friedman.
\newblock Introducing github copilot: your {AI} pair programmer, 2021.

\bibitem[\protect\citeauthoryear{Handsaker}{1982}]{MITcodegeneration}
Robert~E. Handsaker.
\newblock Code generation in the programmer's apprentice.
\newblock Working Paper 233, {MIT} {AI} Lab, May 1982.

\bibitem[\protect\citeauthoryear{Hindle \bgroup \em et al.\egroup
  }{2012}]{hindle2012naturalness}
Abram Hindle, Earl~T Barr, Zhendong Su, Mark Gabel, and Premkumar Devanbu.
\newblock On the naturalness of software.
\newblock In {\em 2012 34th International Conference on Software Engineering},
  pages 837--847. IEEE, 2012.

\bibitem[\protect\citeauthoryear{Husain \bgroup \em et al.\egroup
  }{2019}]{husain2019codesearchnet}
Hamel Husain, Ho-Hsiang Wu, Tiferet Gazit, Miltiadis Allamanis, and Marc
  Brockschmidt.
\newblock {CodeSearchNet} challenge: Evaluating the state of semantic code
  search.
\newblock {\em arXiv preprint arXiv:1909.09436}, 2019.

\bibitem[\protect\citeauthoryear{Imai}{2022}]{imai2022github}
Saki Imai.
\newblock Is github copilot a substitute for human pair-programming? an
  empirical study.
\newblock In {\em Proceedings of the ACM/IEEE 44th International Conference on
  Software Engineering: Companion Proceedings}, pages 319--321, 2022.

\bibitem[\protect\citeauthoryear{Ji \bgroup \em et al.\egroup
  }{2020}]{ji2020amazing}
Yangfeng Ji, Antoine Bosselut, Thomas Wolf, and Asli Celikyilmaz.
\newblock The amazing world of neural language generation.
\newblock In {\em Proceedings of the 2020 Conference on Empirical Methods in
  Natural Language Processing: Tutorial Abstracts}, pages 37--42, 2020.

\bibitem[\protect\citeauthoryear{Kontogiannis \bgroup \em et al.\egroup
  }{1996}]{kontogiannis1996pattern}
Kostas~A Kontogiannis, Renator DeMori, Ettore Merlo, Michael Galler, and Morris
  Bernstein.
\newblock Pattern matching for clone and concept detection.
\newblock {\em Automated Software Engineering}, 3(1-2):77--108, 1996.

\bibitem[\protect\citeauthoryear{Li \bgroup \em et al.\egroup
  }{2022a}]{li2022competition}
Yujia Li, David Choi, Junyoung Chung, Nate Kushman, Julian Schrittwieser,
  R{\'e}mi Leblond, Tom Eccles, James Keeling, Felix Gimeno, Agustin Dal~Lago,
  et~al.
\newblock Competition-level code generation with alphacode.
\newblock {\em Science}, 378(6624):1092--1097, 2022.

\bibitem[\protect\citeauthoryear{Li \bgroup \em et al.\egroup
  }{2022b}]{li2022automating}
Zhiyu Li, Shuai Lu, Daya Guo, Nan Duan, Shailesh Jannu, Grant Jenks, Deep
  Majumder, Jared Green, Alexey Svyatkovskiy, Shengyu Fu, et~al.
\newblock Automating code review activities by large-scale pre-training.
\newblock In {\em Proceedings of the 30th ACM Joint European Software
  Engineering Conference and Symposium on the Foundations of Software
  Engineering}, 2022.

\bibitem[\protect\citeauthoryear{Manna and Waldinger}{1971}]{manna1971toward}
Zohar Manna and Richard~J Waldinger.
\newblock Toward automatic program synthesis.
\newblock {\em Communications of the ACM}, 14(3):151--165, 1971.

\bibitem[\protect\citeauthoryear{Mozannar \bgroup \em et al.\egroup
  }{2022}]{mozannar2022reading}
Hussein Mozannar, Gagan Bansal, Adam Fourney, and Eric Horvitz.
\newblock Reading between the lines: Modeling user behavior and costs in
  {AI}-assisted programming.
\newblock {\em arXiv preprint arXiv:2210.14306}, 2022.

\bibitem[\protect\citeauthoryear{OpenAI}{2023a}]{openai_2023}
OpenAI.
\newblock Chatgpt: Optimizing language models for dialogue, Jan 2023.

\bibitem[\protect\citeauthoryear{OpenAI}{2023b}]{gpt42023}
OpenAI.
\newblock Gpt-4 technical report.
\newblock {\em arXiv preprint arXiv:2303.08774}, 2023.

\bibitem[\protect\citeauthoryear{Ouyang \bgroup \em et al.\egroup
  }{2022}]{ouyang2022training}
Long Ouyang, Jeffrey Wu, Xu~Jiang, Diogo Almeida, Carroll Wainwright, Pamela
  Mishkin, Chong Zhang, Sandhini Agarwal, Katarina Slama, Alex Ray, et~al.
\newblock Training language models to follow instructions with human feedback.
\newblock {\em Advances in Neural Information Processing Systems},
  35:27730--27744, 2022.

\bibitem[\protect\citeauthoryear{Poldrack \bgroup \em et al.\egroup
  }{2023}]{poldrack2023ai}
Russell~A Poldrack, Thomas Lu, and Ga{\v{s}}per Begu{\v{s}}.
\newblock {AI}-assisted coding: Experiments with gpt-4.
\newblock {\em arXiv preprint arXiv:2304.13187}, 2023.

\bibitem[\protect\citeauthoryear{Raffel \bgroup \em et al.\egroup
  }{2020}]{raffel2020exploring}
Colin Raffel, Noam Shazeer, Adam Roberts, Katherine Lee, Sharan Narang, Michael
  Matena, Yanqi Zhou, Wei Li, and Peter~J Liu.
\newblock Exploring the limits of transfer learning with a unified text-to-text
  transformer.
\newblock {\em The Journal of Machine Learning Research}, 2020.

\bibitem[\protect\citeauthoryear{Rajamani}{2022}]{rajamani2022ai}
Sriram Rajamani.
\newblock {AI} assisted programming.
\newblock In {\em 15th Annual ACM India Compute Conference}, COMPUTE '22,
  page~5, New York, NY, USA, 2022. Association for Computing Machinery.

\bibitem[\protect\citeauthoryear{Rich and Waters}{1982}]{MITsimplify}
Charles Rich and Richard~C. Waters.
\newblock The disciplined use of simplifying assumptions.
\newblock {\em ACM SIGSOFT Software Engineering Notes}, 7(5):150--154, December
  1982.

\bibitem[\protect\citeauthoryear{Rich and Waters}{1988}]{MITapprentice88}
Charles Rich and Richard~C. Waters.
\newblock The programmer's apprentice: a research overview.
\newblock {\em Computer}, 21(11):10--25, November 1988.

\bibitem[\protect\citeauthoryear{Rich \bgroup \em et al.\egroup
  }{1978}]{MITprompt}
Charles Rich, Howard~E. Shrobe, Robert~C. Waters, Gerald~J. Sussman, and
  Carl~E. Hewitt.
\newblock Programming viewed as an engineering activity.
\newblock {AI} Memo 459, Massachusetts Institute of Technology, January 1978.

\bibitem[\protect\citeauthoryear{Robbes and Lanza}{2008}]{robbes2008program}
Romain Robbes and Michele Lanza.
\newblock How program history can improve code completion.
\newblock In {\em 23rd IEEE/ACM International Conference on Automated Software
  Engineering}, pages 317--326, 2008.

\bibitem[\protect\citeauthoryear{Saha \bgroup \em et al.\egroup
  }{2017}]{saha2017elixir}
Ripon~K Saha, Yingjun Lyu, Hiroaki Yoshida, and Mukul~R Prasad.
\newblock Elixir: Effective object-oriented program repair.
\newblock In {\em 2017 32nd IEEE/ACM International Conference on Automated
  Software Engineering}, pages 648--659. IEEE, 2017.

\bibitem[\protect\citeauthoryear{Siracusa \bgroup \em et al.\egroup
  }{2023}]{applepatent}
M.~R. Siracusa, A.~K. Katti, and et~al.
\newblock Integrating learning models into software development systems, June
  27th 2023.
\newblock US Patent and Trademark Office, US Patent 11,687,830.

\bibitem[\protect\citeauthoryear{Sridhara \bgroup \em et al.\egroup
  }{2010}]{sridhara2010towards}
Giriprasad Sridhara, Emily Hill, Divya Muppaneni, Lori Pollock, and
  K~Vijay-Shanker.
\newblock Towards automatically generating summary comments for java methods.
\newblock In {\em IEEE/ACM International Conference on Automated Software
  Engineering}, pages 43--52, 2010.

\bibitem[\protect\citeauthoryear{Sridhara \bgroup \em et al.\egroup
  }{2011}]{sridhara2011generating}
Giriprasad Sridhara, Lori Pollock, and K~Vijay-Shanker.
\newblock Generating parameter comments and integrating with method summaries.
\newblock In {\em 2011 IEEE 19th International Conference on Program
  Comprehension}, pages 71--80. IEEE, 2011.

\bibitem[\protect\citeauthoryear{Surameery and Shakor}{2023}]{surameery2023use}
Nigar M~Shafiq Surameery and Mohammed~Y Shakor.
\newblock Use chatgpt to solve programming bugs.
\newblock {\em International Journal of Information Technology \& Computer
  Engineering (IJITC) ISSN: 2455-5290}, 3(01):17--22, 2023.

\bibitem[\protect\citeauthoryear{Talamadupula}{2021}]{talamadupula2021applied}
Kartik Talamadupula.
\newblock Applied {AI} matters: Ai4code: Applying artificial intelligence to
  source code.
\newblock {\em {AI} Matters}, 7(1):18--20, 2021.

\bibitem[\protect\citeauthoryear{Vaswani \bgroup \em et al.\egroup
  }{2017}]{vaswani2017attention}
Ashish Vaswani, Noam Shazeer, Niki Parmar, Jakob Uszkoreit, Llion Jones,
  Aidan~N Gomez, {\L}ukasz Kaiser, and Illia Polosukhin.
\newblock Attention is all you need.
\newblock {\em Advances in Neural Information Processing Systems}, 2017.

\bibitem[\protect\citeauthoryear{Vechev \bgroup \em et al.\egroup
  }{2016}]{vechev2016programming}
Martin Vechev, Eran Yahav, et~al.
\newblock Programming with “big code”.
\newblock {\em Foundations and Trends{\textregistered} in Programming
  Languages}, 3(4):231--284, 2016.

\bibitem[\protect\citeauthoryear{Waldinger and Lee}{1969}]{waldinger1969prow}
Richard~J Waldinger and Richard~CT Lee.
\newblock Prow: A step toward automatic program writing.
\newblock In {\em 1st International Joint Conference on Artificial
  Intelligence}, pages 241--252, 1969.

\bibitem[\protect\citeauthoryear{Waters}{1982}]{MITapprentice82}
Richard~C. Waters.
\newblock The programmer's apprentice: Knowledge based program editing.
\newblock {\em IEEE Transactions on Software Engineering}, SE-8(1):1--12,
  January 1982.

\bibitem[\protect\citeauthoryear{Wong \bgroup \em et al.\egroup
  }{2023}]{wong2023natural}
Man-Fai Wong, Shangxin Guo, Ching-Nam Hang, Siu-Wai Ho, and Chee-Wei Tan.
\newblock Natural language generation and understanding of big code for
  {AI}-assisted programming: A review.
\newblock {\em Entropy}, 25(6):888, 2023.

\bibitem[\protect\citeauthoryear{Wu \bgroup \em et al.\egroup
  }{2022}]{wu2022promptchainer}
Tongshuang Wu, Ellen Jiang, Aaron Donsbach, Jeff Gray, Alejandra Molina,
  Michael Terry, and Carrie~J Cai.
\newblock Promptchainer: Chaining large language model prompts through visual
  programming.
\newblock In {\em CHI Conference on Human Factors in Computing Systems Extended
  Abstracts}, pages 1--10, 2022.

\bibitem[\protect\citeauthoryear{Zheng \bgroup \em et al.\egroup
  }{2015}]{zheng2015bid}
Liang Zheng, Carlee Joe-Wong, Chee~Wei Tan, Mung Chiang, and Xinyu Wang.
\newblock How to bid the cloud.
\newblock In {\em Proceedings of the 2015 ACM Conference on Special Interest
  Group on Data Communication (SIGCOMM)}, pages 71--84, 2015.

\bibitem[\protect\citeauthoryear{Zheng \bgroup \em et al.\egroup
  }{2016}]{zheng2016viability}
Liang Zheng, Carlee Joe-Wong, Christopher~G Brinton, Chee~Wei Tan, Sangtae Ha,
  and Mung Chiang.
\newblock On the viability of a cloud virtual service provider.
\newblock In {\em Proceedings of the 2016 ACM SIGMETRICS International
  Conference on Measurement and Modeling of Computer Science}, pages 235--248,
  2016.

\end{thebibliography}
\end{document}